\documentclass[preprint2]{aastex}
\usepackage{graphicx}
\usepackage{amsmath}
\usepackage[caption=false]{subfig}
\usepackage{natbib}
\usepackage{booktabs}
\usepackage{lineno}

\begin{document}

\title{The Confinement of the Heliosheath Plasma by the Solar Magnetic Field 
as Revealed by Energetic Neutral Atom Simulations}

\author{M. Kornbleuth\altaffilmark{1}, M. Opher\altaffilmark{1}, A. T. Michael\altaffilmark{1}$^{,}$\altaffilmark{2}, J. M. Sok{\'o}{\l}\altaffilmark{3}, G. T{\'o}th\altaffilmark{4}, V. Tenishev\altaffilmark{4}, J. F. Drake\altaffilmark{5}}
\affil{\altaffilmark{1}Astronomy Department, Boston University, Boston, MA 
02215, USA} 
\email{kmarc@bu.edu}

\affil{\altaffilmark{2}Applied Physics Laboratory, Johns Hopkins University, 
Laurel, MD 20723, USA}

\affil{\altaffilmark{3}NAWA Bekker Fellow, Department of Astrophysical Sciences, Princeton University,
Princeton, NJ 08544, USA}

\affil{\altaffilmark{4}University of Michigan, Ann Arbor, MI 48109, USA}

\affil{\altaffilmark{5}Department of Physics and the Institute for Physical 
Science and Technology, University of Maryland, College Park, MD, USA}

\begin{abstract}
Traditionally, the solar magnetic field has been considered to have a 
negligible effect in the outer regions of the heliosphere. Recent works 
have shown that the solar magnetic field may play a crucial role in collimating the plasma in the 
heliosheath. \textit{Interstellar Boundary Explorer} (\textit{IBEX}) 
observations of the heliotail indicated a latitudinal structure varying with energy in the energetic 
neutral atom (ENA) fluxes. At energies $\sim$1 keV, the ENA fluxes show an 
enhancement at low latitudes and a deficit of ENAs near the poles. At energies $>$2.7 keV, ENA fluxes 
had a deficit within low latitudes, and lobes of higher ENA flux near the poles. This ENA structure 
was initially interpreted to be a result of the latitudinal profile of the solar wind 
during solar minimum. We extend the work of \citet{Kornbleuth18} by using solar minimum-like 
conditions and the recently developed SHIELD model. The SHIELD model couples the 
magnetohydrodynamic (MHD) plasma 
solution with a kinetic description of neutral hydrogen. We show that while the latitudinal 
profile of the solar wind during solar minimum contributes to the lobes in ENA maps, the 
collimation by the solar magnetic field is important in creating and shaping the two high 
latitude lobes of enhanced ENA flux observed by \textit{IBEX}. This is the first work to explore the 
effect of the changing solar magnetic field strength on ENA maps. Our findings suggest that 
\textit{IBEX} is providing the first observational evidence of the collimation of the heliosheath 
plasma by the solar magnetic field.
\end{abstract}


\section{Introduction}

The heliosphere is created via the interaction of the solar wind plasma with 
the partially ionized plasma of the local interstellar medium (ISM). The 
solar cycle consists of a progression from solar minimum ($\sim$400 
km/s solar wind at low latitudes, $\sim$750 km/s wind at high latitudes) to 
solar maximum ($\sim$400 km/s solar wind at all latitudes) and back to solar 
minimum. The solar wind dynamic pressure varies as the solar wind speed 
and density changes in time affecting the heliospheric boundaries and the 
energetic neutral atom (ENA) production \citep{McComas18, McComas19, Zirnstein18}. 
Additionally, the solar magnetic field intensity changes in magnitude during the 
course of the solar cycle.

Traditionally, the accepted shape of the heliosphere is comet-like \citep{Parker61, Baranov93}, with a
long tail extending for thousands of AU in the direction opposite ISM flow (the heliotail). 
Recently, this notion was challenged by models and observations. \citet{Opher15} 
used a magnetohydrodynamic (MHD) simulation to propose that the magnetic tension 
of the solar magnetic field alters flows in the heliosheath to produce a 
``croissant" heliosphere with lobes directed to the north and south. This 
model includes a shortened heliotail, where the distance from the Sun to 
the tail is similar to the distance from the Sun to the nose. \citet{Dialynas17} proposed a 
bubble-like heliosphere, also with a shortened heliotail, based on ENA measurements (30-55 keV) from 
the Ion and Neutral Camera onboard \textit{Cassini}. One important feature of the \citet{Opher15} 
model was the ability of the solar magnetic field to collimate the solar wind plasma. 
\citet{Pogorelov15} and \citet{Izmodenov15,Izmodenov18} argue for a long-tail heliosphere, but show evidence of 
this collimation process while using a unipolar magnetic field. 

Energetic Neutral Atoms (ENAs) originate as pick-up ions (PUIs) in the solar wind plasma and 
can provide an indirect method for studying the structural properties of the heliosphere. The 
\textit{Interstellar Boundary Explorer} (\textit{IBEX}) was launched 2008 October 19 
to study the heliosphere by observing ENAs \citep{McComas09a}. The \textit{IBEX-Hi} camera 
\citep{Funsten09} measures ENAs from $\sim$ 300 eV to $\sim$ 6 keV. ENA images of the heliotail by 
\textit{IBEX} \citep{McComas13} show a multi-lobe structure. These lobes were attributed to 
signatures of slow and fast wind within the long heliospheric tail as part of the 
11-year solar cycle. \citet{Zirnstein16a,Zirnstein17}, using a time dependent 
simulation with solar cycle variations of the solar wind, reproduced the spectral 
dependence and heliotail ENA maps as shown by \textit{IBEX} with a model of the heliotail 
resembling a long comet-like tail. 

\citet{Kornbleuth18} investigated the ENA maps from the ``croissant" heliosphere 
\citep{Opher15} with a uniform solar wind. The collimation of the heliosheath plasma 
by the solar magnetic field created high latitude lobes despite the lack of fast solar wind
at the poles. These high latitude lobes persisted in all energies contrary to observations. 
\citet{Kornbleuth18} included a solar magnetic field with a radial component of 6.45 nT at 1 AU, 
representative of solar maximum-like conditions as in \citet{Opher15}, which is higher than other models such as 3.5 nT in \citet{Pogorelov15} and 3.75 nT in 
\citet{Izmodenov15}. These lower values for the radial solar magnetic field 
component reflect the field strength seen at solar minimum. 

By using a time dependent MHD solution as in \citet{Zirnstein17}, it is difficult 
to disentangle the effects of a latitudinally dependent solar wind and the solar 
magnetic field. Previous studies suggested the latitudinally dependent solar wind 
is responsible for the energy dependent ENA structures in the heliotail observed 
by \textit{IBEX} \citep{McComas13, Zirnstein16a, Zirnstein17}. \citet{Kornbleuth18} 
suggested the solar magnetic field could play a key role in the formation of 
observed high latitude lobes, but only used solar maximum-like conditions. Therefore, it is critical 
to disentangle the effects of the solar wind structure and the solar magnetic field to understand the 
origin of ENAs in observations.

The purpose of this paper is to extend our previous work \citep{Kornbleuth18} to investigate the 
effect of a latitudinally-varying solar wind and the strength of the solar magnetic 
field. Other works, such as \citet{Ratkiewicz12}, have studied the effect of the interstellar 
magnetic field on ENA maps. In contrast, the work in this paper focuses on the effect of the solar 
magnetic field, which has often been regarded as having a negligible effect. We use conditions from the
year 2008, corresponding to a period of solar minimum. We compare with the first five years of 
\textit{IBEX} data that correspond to the solar minimum period. In Section 2, we discuss the model used 
in this work. In Section 3, we present the results of our model, focusing on the different roles of the 
solar wind structure and the solar magnetic field. In Section 4, we discuss the implications of our 
modeling.

\section{Model}

\subsection{MHD Simulations} \label{ssec:MHD}

\begin{figure*}[t!]
\centering
  \includegraphics[scale=0.45]{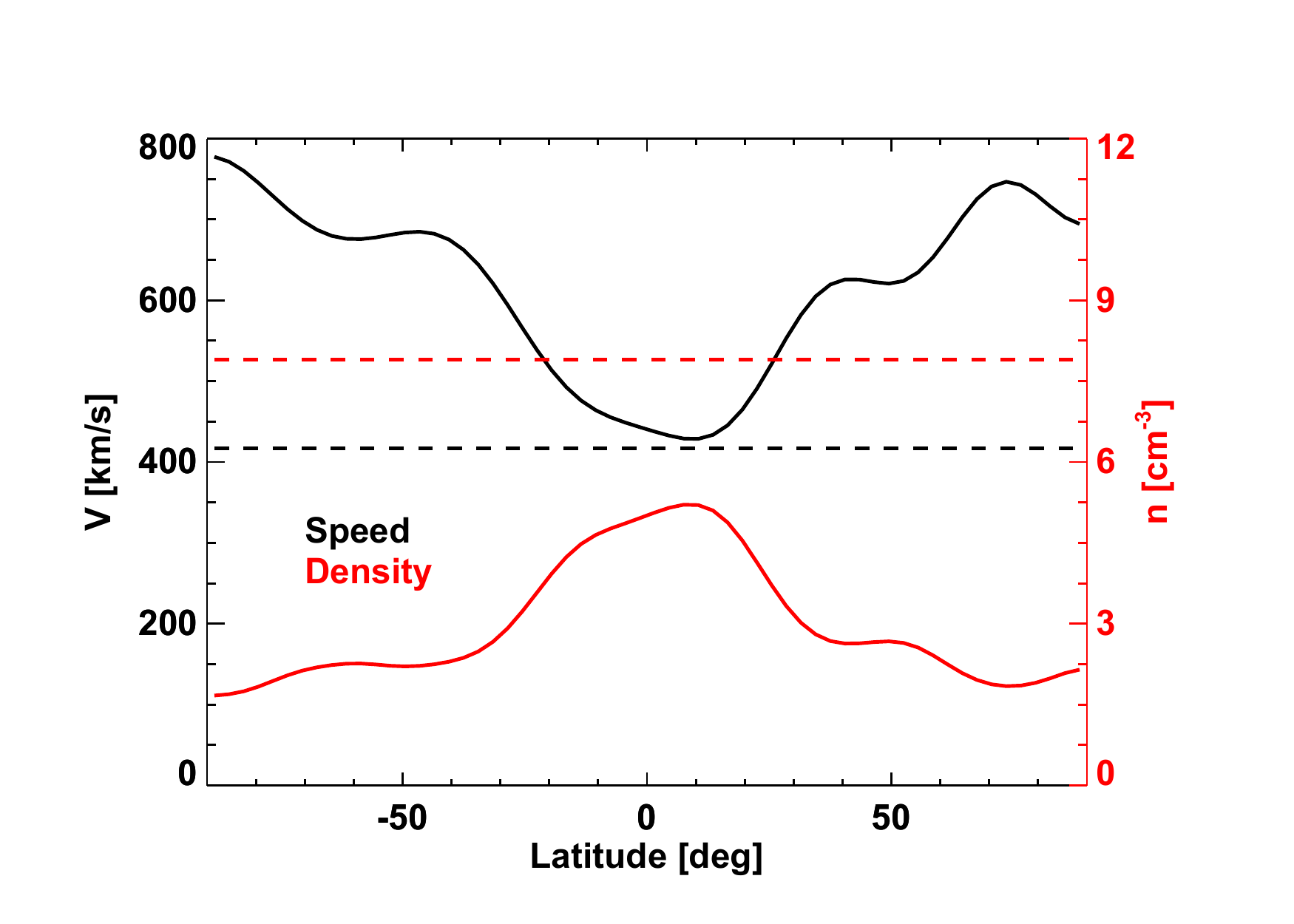}
  \includegraphics[scale=0.45]{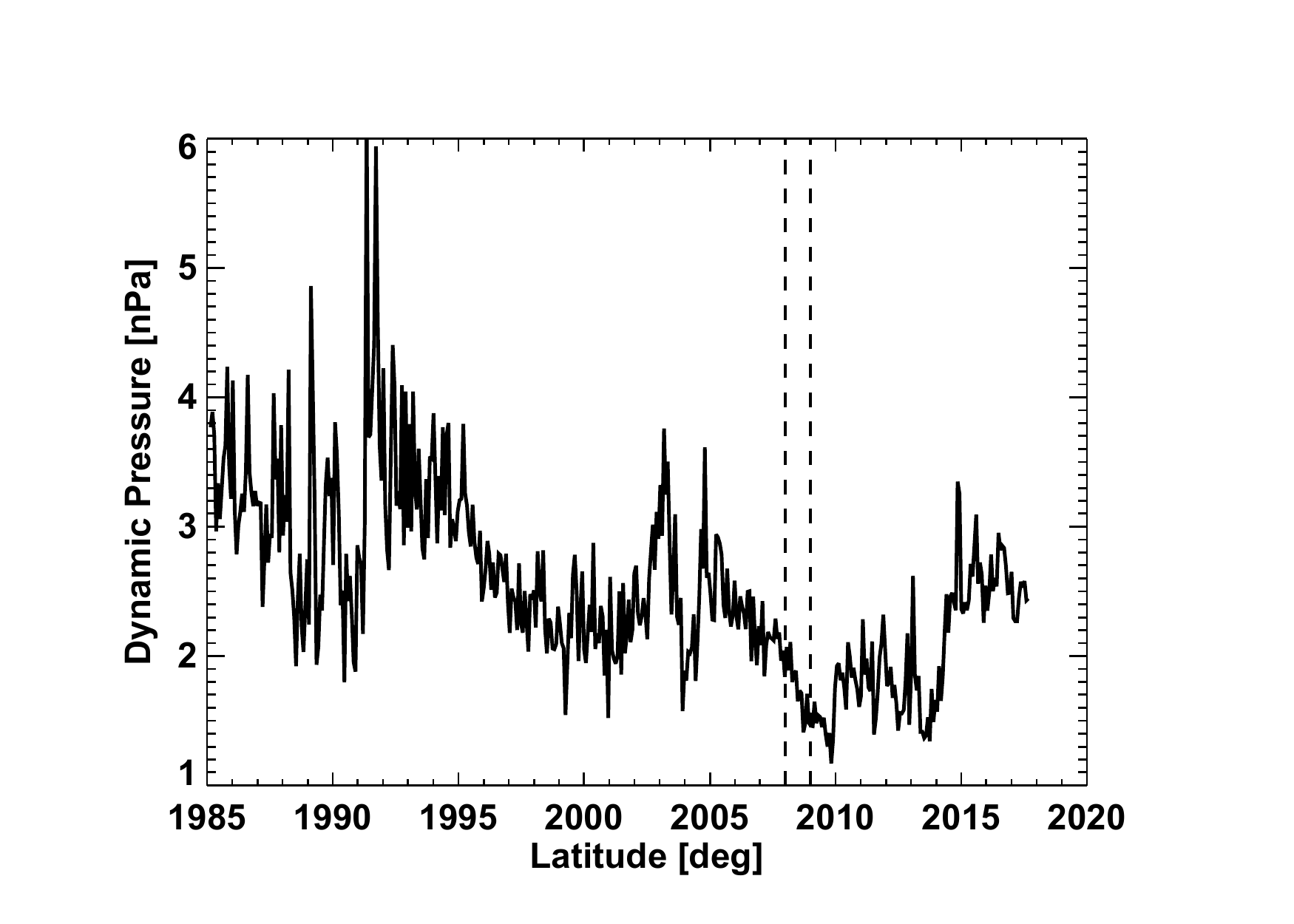}
  \caption{Left: Solar wind conditions at 1 AU for different heliolatitudes corresponding to the 2008-averaged solar wind data from \citet{Sokol15} used in Cases 1 and 2 (solid) and the uniform solar wind used in Cases 3 and 4 (dashed). The black lines correspond to the solar wind speed and the red lines correspond to the solar wind density. Right: Variation of solar wind dynamic pressure with time for the equator at 1 AU from \citet{Sokol15}. The dashed vertical lines indicate the start and end of the year 2008, over which the solar wind data in Cases 1 and 2 are averaged.}
  \label{ics}
\end{figure*}

We use the Solar-wind with Hydrogen Ion Exchange and Large-scale Dynamics (SHIELD) model from 
\citet{Michael19} and \citet{Michael20}. This model extends the MHD solution of \citet{Opher15} to 
become a global, self-consistent kinetic-MHD model of the heliosphere.
The model treats the plasma as a single ion species using the MHD equations and describes the neutral 
hydrogen kinetically. Our model uses a Cartesian grid, with the outer boundary 
located at $x=\pm1500$ AU, $y=\pm2000$ AU, and $z=\pm2000$ AU. The neutrals flow 
throughout the domain, and the ISM enters the domain from the $x=-1500$ AU face. 
For the ISM, we use the following parameters from \citet{Opher20}: $n_{H}=0.18$ 
cm$^{-3}$ for the neutral hydrogen, $n_{p}=0.06$ cm$^{-3}$ for the ISM protons, $v_{ISM}
=26.4$ km s$^{-1}$, $T_{ISM}=6519$ K, and $B_{ISM}=3.2$ $\mu$G. These boundaries 
are chosen because they give good agreement with the measurements of Voyager 1 \& 
2 outside of the heliopause.

Our inner boundary conditions are specified at 10 AU. We use the solar wind speed
and density from \citet{Sokol15}. These authors used the solar wind speed derived
from the interplanetary scintillation (IPS) observations by the Institute for 
Space-Earth Environmental Research at Nagoya University \citep{Tokumaru10, Tokumaru12}.
They reconstructed the spatial and temporal variation of the solar wind speed and density
at 1 AU from 1985 to 2013. We use yearly-averaged solar wind data corresponding to the year 2008 for 
our inner boundary conditions with 3$^{\circ}$ resolution in latitude and assume longitudinal 
symmetry (Figure \ref{ics}). This year corresponds to typical solar minima conditions. We determine 
the solar wind temperature at 1 AU using those data and assume a Mach number of 6 at all latitudes 
at 1 AU. We extrapolate the solar wind conditions to 10 AU assuming the flow to be adiabatic. As in 
\citet{Opher15}, we treat the solar magnetic field as a unipole with the same polarity in both 
hemispheres to avoid artificial numerical reconnection at both the nose of the heliosphere and also in
the solar magnetic equator across the heliospheric current sheet. The radial component of the solar 
magnetic field is set at the equator as 2.94 nT  at 1 AU. This value was taken from yearly-averaged 
OMNI magnetic field data for the year 2008 at 1 AU. The solar magnetic field is modeled as a Parker 
spiral \citep{Parker58}.

\begin{table*}[t!]
\centering
\begin{tabular}{ccccccccc}
\tableline
Case && n cm$^{-3}$ && U km/s & & B$_{r}$ nT & & Latitudinal Variation\\
&& && && && of Solar Wind\\
\tableline
Case 1 & & 5.1 && 437.7 && 0 & & yes\\
Case 2 & & 5.1 && 437.7 && 2.94 & & yes \\
Case 3 & & 7.9 && 417.1 && 2.94 & & no  \\
Case 4 & & 7.9 && 417.1 && 6.45 & & no \\
\tableline
\end{tabular}
\caption{The four cases used in this work. The solar wind density (n) and speed (U), 
as well as the radial component of the solar magnetic field (B$_{r}$) are given for each 
case at 1 AU for the solar equator. Cases 1 and 2 represent solar 
minimum-like conditions with slow wind at the low latitudes and fast wind at the high latitudes. 
Cases 3 and 4 represent solar maximum-like conditions, with a uniform solar wind profile 
across all latitudes. For all cases, longitudinal symmetry is assumed.}
\label{bcs}
\end{table*}

The SHIELD model uses the Adaptive Mesh Particle Simulator (AMPS) to treat 
the neutral hydrogen kinetically. AMPS is a global, kinetic, 3D particle code
developed within the framework of the Direct Simulation Monte Carlo methods 
\citep{Tenishev08}. The multi-fluid approximation, which describes the neutrals as four 
separate fluids, is used to relax the plasma to a steady state solution in the MHD model, and is
used to start SHIELD. The resolution of the MHD model is ${\Delta}x= 3$ AU in the supersonic solar 
wind (within $x={\pm}120$ AU) and ${\Delta}x= 6$ AU in the heliosheath (from $x=-240$ AU 
at the nose until $x=560$ AU at the heliotail).  The resolution of AMPS within the heliosphere is
${\Delta}x= 4.7$ AU (from $x=-280$ AU at the nose until $x=560$ AU at the heliotail).The SHIELD model 
is run in local time to cycle between the MHD model and AMPS, which allows for statistics to 
accumulate between each step of the plasma solution \citep{Michael20}. Use of a local time step 
\citep{Toth12} is appropriate for steady state solutions. Approximately 140 million particles are 
modeled within AMPS, and the source terms to the MHD equations due to charge-exchange with the neutral
hydrogen atoms accrue for 5,000 time steps within AMPS before being passed to the MHD solver to update
the plasma solution for one time step. Once the plasma solution is updated, it is passed back to AMPS.
This process is repeated until a new approximate steady state solution is reached.

In order to probe the effects of a latitudinally-varying solar wind and the strength
of the solar magnetic field on the solution, we run four different cases (Table 
\ref{bcs}). Cases 1 and 2 correspond to solar minimum-like conditions, while Cases
3 and 4 correspond to solar maximum-like conditions. Case 1 utilizes
the yearly-averaged 2008 solar wind conditions (Figure \ref{ics}) from \citet{Sokol15}; however, the 
solar magnetic field is not included. Case 2 includes the 2008 solar wind conditions and also the 
solar magnetic field. Case 3 is based on the model used in \citet{Kornbleuth18}. We use the outer 
boundary conditions detailed above and the same solar magnetic field as in Cases 1 and 2, but we use a
uniform solar wind profile  (as in \citealt{Kornbleuth18}) corresponding to solar maximum-like 
conditions from \citep{Opher15} with a density of $n_{SW}=7.9$ cm$^{-3}$, temperature of $T_{SW}=2.9 
\times 10^{5}$ K, and speed of $v_{SW}=417.1$ km/s at 1 AU (Figure \ref{ics}). Case 4 utilizes the 
same conditions as Case 3, except the radial solar magnetic field at the equator is set to be 6.45 nT 
at 1 AU instead of 2.94 nT. This intensity corresponds to solar maximum-like conditions used in 
\citet{Opher15}.

\subsection{ENA Flux Model} \label{ssec:ENA}

We update the model of \citet{Kornbleuth18} to account for variation in solar wind speed and density 
at different latitudes in the supersonic solar wind. When including the latitudinal variation of the 
solar wind, the speed and density of the plasma must be considered with regard to the creation of 
PUIs via charge exchange.

We use the PUI density fraction at the termination shock, $\alpha$, from 
\citet{Zirnstein17}, which is based on the work of \citet{Lee09}, given by

\begin{equation}
\begin{split}
\alpha(\mathbf{r_{TS}})=\frac{r_{TS}}{u_{p,1 au}n_{p,1 au}}n_{H,avg}
(\nu_{ph,1 au}\\
+\sigma_{ex}u_{p,1 au}n_{p,1 au}),
\end{split}
\label{eq:puirat}
\end{equation}

\noindent where $r_{TS}$ is the distance for the termination shock, $u_{p,1 
au}$ is the speed of the solar wind at 1 AU, $n_{p,1 au}$ is the density 
of the solar wind at 1 AU, and $n_{H,avg}$ is the average neutral hydrogen density
between the inner boundary and the termination shock for a given direction. The term 
$\nu_{ph,1 au}$ is the photoionization rate at 1 AU assumed to be $8\times10^{-8}$ s$^{-1}$ 
(e.g., \citealt{Sokol19}). As in \citet{Zirnstein17}, we assume the constant photoionization rate, 
although it varies in time and space. The photoionization rate contributes about 20$\%$ to 
the total ionization rates for hydrogen, with charge exchange being the dominant ionization process (see more, e.g., \citealt{Sokol19}). Therefore our approximation does not change the conclusions. The 
charge-exchange cross section, $\sigma_{ex}$, is used from \citet{Lindsay05}.

Similar to \citet{Kornbleuth18}, we include the angular dependence of PUIs in our 
model via

\begin{equation}
n_{PUI}(\mathbf{r})=n_{i}(\mathbf{r})\frac{\alpha(\theta,\phi)}
{\alpha(\theta_{nose},\phi_{nose})},
\label{eq:ang_pui}
\end{equation}

\noindent where $\theta$ is the polar angle, and $\phi$ is the 
azimuthal angle. The polar angle increases from the northern 
pole toward the southern pole, while the azimuthal angle increases in the 
clockwise direction from the tail. The parameter $n_{i}$ is the 
density of the PUI species for a given direction calculated using the 
total plasma density multiplied by the density fractions \citep{Kornbleuth18}. We 
partition the plasma from the MHD solution into three populations of ions based on the work of 
\citet{Zank10}: solar wind ions, transmitted PUIs, and reflected PUIs, with their 
density ratios relative to the plasma given by 0.836, 0.151, and 0.013, respectively, 
and their energy ratios given by 0.04, 0.50, and 0.46, respectively. The total thermal 
energy of the plasma is conserved via this partitioning, given by \citep{Zank10}

\begin{equation}
T_{p}=\left(\frac{n_{\mathrm{SW}}}{n_{p}}\Gamma_{\mathrm{SW}}+
\frac{n_{\mathrm{tr}}}{n_{p}}\Gamma_{\mathrm{tr}}+\frac{n_{\mathrm{ref}}}
{n_{p}}\Gamma_{\mathrm{ref}}\right)T_{p},
\label{eq:Zank}
\end{equation}

\noindent where $n_{p}$ and $T_{p}$ are the density and temperature of the 
plasma, $n_{i}$ is the density for the respective ion population, and $\Gamma_{i}$ is 
the temperature fraction for the respective ion population given by $\Gamma_{i}=T_{i}/T_{p}$, with 
$T_{i}$ being the temperature for the respective ion population. We use the cold electron 
approximation for the plasma, where the electrons are assumed to have the same 
temperature as the solar wind ions. Therefore, the plasma temperature is given by 
\citep{Zirnstein17}

\begin{equation}
T_{p}=\frac{2T_{\mathrm{MHD}}}{1+\Gamma_{SW}},
\end{equation}

\noindent where $T_{\mathrm{MHD}}$ is the temperature given from the MHD 
solution. We also assume quasi-neutrality.

We are not modeling the ENA the contribution from the region in the 
heliotail where material with heliosheath properties sits on magnetic field lines open to the ISM
\citep{Michael18}, called ``open heliosheath". \citet{Kornbleuth18} showed that the ``open 
heliosheath" region could enhance the ENA flux observed in the low latitude tail. Since the purpose of
this work is to focus on the effects of the solar magnetic field and the latitudinal structure 
of the solar wind, the exploration of the contribution of ``open heliosheath" is left to a future work. 

\section{Results}

\begin{figure*}[t!]
\centering
  \includegraphics[scale=0.55]{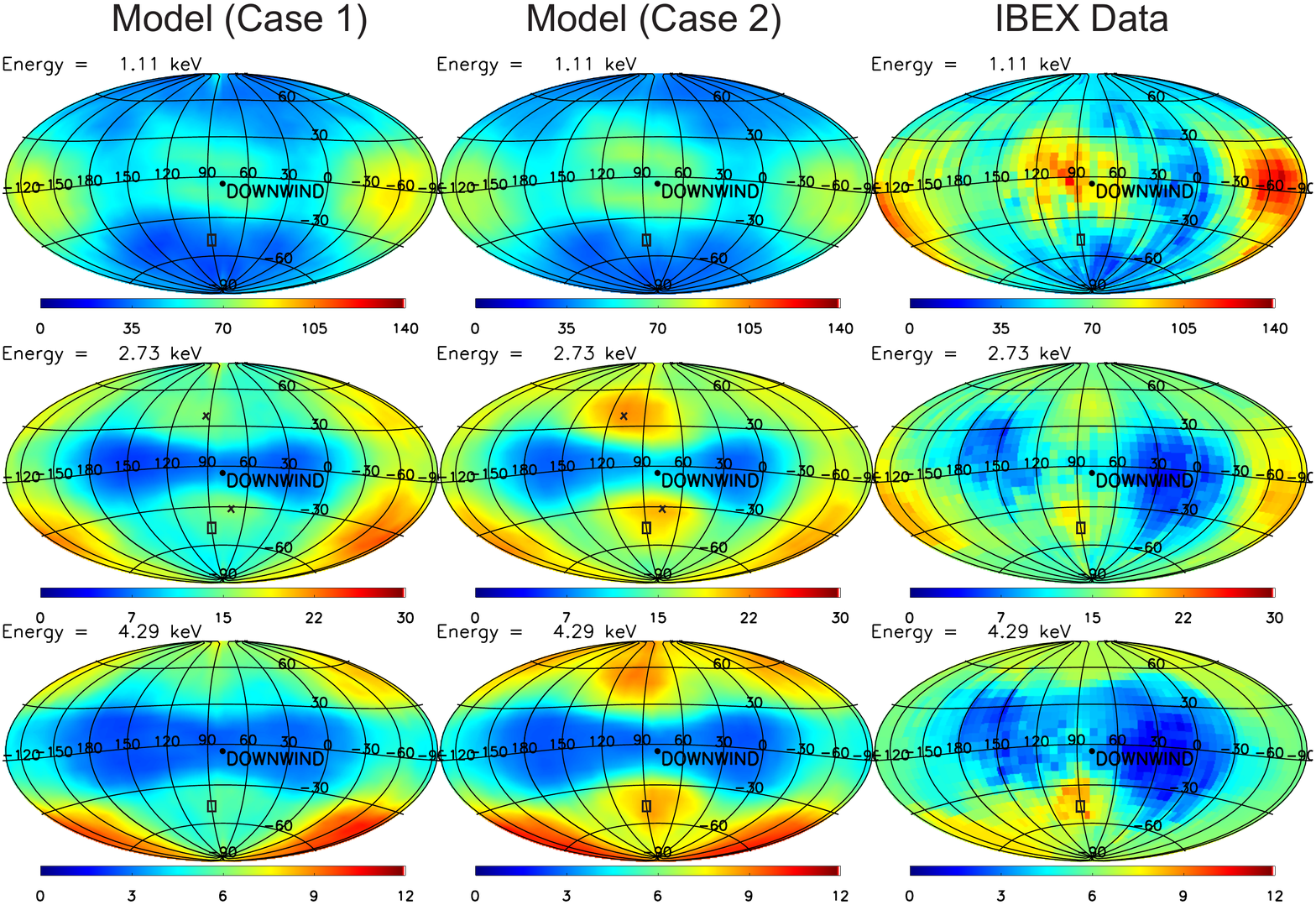}
  \caption{ENA sky map of flux centered on the downwind (tail) direction in units 
  of (cm$^{2}$ s sr keV)$^{-1}$. From top to bottom, the energies 
  included are 1.11, 2.73, and 4.29 keV. Left:  simulated sky maps for Case 1, 	
  which includes latitudinally-varying solar wind without solar magnetic field. 
  Middle: simulated sky maps for Case 2, which includes latitudinally-varying 
  solar wind and solar magnetic field. Right: \textit{IBEX} ENA maps from the first
  five years (2009-2013) of observations \citep{Schwadron14}. Simulated sky maps are 
  multiplied by a factor of 1.8. The black box on each map corresponds to
  the location in the southern lobe where the spectral indices are derived in Table \ref{spec}.
  For Case 1 in the 2.73 keV energy band the maximum flux within the lobes are 
  located at ($\lambda$, $\beta$) = (96$^{\circ}$, 42$^{\circ}$) for the northern lobe 
  and ($\lambda$, $\beta$) = (72$^{\circ}$, -30$^{\circ}$) for the southern lobe. For 
  Case 2, the northern lobe displays maximum flux at ($\lambda$, $\beta$) = (108$^{\circ}$, 
  42$^{\circ}$) and the southern lobe has peak flux centered at ($\lambda$, $\beta$) = 
  (78$^{\circ}$, -30$^{\circ}$) for the 2.73 keV energy band. The location of maximum flux 
  within the lobes of Cases 1 and 2 are marked by a black `x'. }
  \label{tail}
\end{figure*}

In Figure \ref{tail}, we present the tail-centered maps from our different cases, 
and tail-centered maps of \textit{IBEX} data from the first five years 
\citep{Schwadron14}. \citet{Schwadron14} applied a mask over the region 
surrounding the \textit{IBEX} ribbon in order to observe the globally distributed flux 
(GDF) from ENAs within the heliosheath using \textit{IBEX} data averaged over the first 
five years. Similar to \citet{Zirnstein17}, we scale the ENA flux in our 
modeled maps by a certain factor (1.8) to match observations. Case 1, which
neglects the solar magnetic field, demonstrates a stronger lower energy ENA flux at 
low latitudes compared to high latitudes. With increasing energy, the ENA 
flux at high latitudes increases relative to the ENA flux at lower latitudes. 
This transition is a consequence of the latitudinally-dependent solar wind 
profile, where ENA flux at the highest energies of \textit{IBEX} are dominated by 
parent ions from the fast solar wind. At energies $>$ 2 keV, the ENA maps have 
two separated lobes of enhanced ENA flux. The peak flux of the northern 
lobe is centered at an ecliptic (longitude, latitude) ($\lambda$, $\beta$) =
(96$^{\circ}$, 42$^{\circ}$) and the southern lobe has a peak flux located
at ($\lambda$, $\beta$) = (72$^{\circ}$, -30$^{\circ}$) in the 2.73 keV energy band.
There is also the the appearance of port and starboard lobes discussed in \citet{McComas13} and \citet{Zirnstein16a}. 
We see these same features in Case 2, which includes the solar magnetic 
field, with the northern lobe having peak flux centered at ($\lambda$, $\beta$) =
(108$^{\circ}$, 42$^{\circ}$) and the southern lobe having peak flux at
($\lambda$, $\beta$) = (78$^{\circ}$, -30$^{\circ}$) in the 2.73 keV energy band.
The high latitude lobes at the 2.73 and 4.29 keV energy bands are significantly enhanced 
in size and flux relative to the case without solar magnetic field. The collimation of the solar wind plasma 
by the solar magnetic field leads to a temperature enhancement of the plasma. 
This temperature increase results in a higher ENA flux because higher temperature
increases the number of parent ions at higher energies. Therefore, the collimation enhances the number 
of higher energy parent ions capable of becoming observed ENAs.

Case 2 displays more ENA flux in the high latitude lobes than both Case 1 and the 
\textit{IBEX} data. Cases 1 and 2 use the same solar wind conditions at the inner boundary
corresponding to the solar minimum and differ only by the inclusion of the solar
magnetic field in Case 2. The results replicate a heliosphere that 
has only experienced solar minimum conditions. \citet{Zirnstein17} showed the 
importance of a time dependent solar wind in ENA modeling. Within the heliosheath at 
high latitudes, there is a mixture of fast and slow solar wind due to the 
progression of the solar cycle. The structure of the heliosheath and the deflection of slow 
solar wind from the low latitudes towards the high latitudes also allows for a mixture of slow and 
fast solar wind at high latitudes independent of the solar cycle. Beyond the deflection of slow 
wind from lower latitudes \citep{Reisenfeld12, Siewert14,Dayeh14}, there is additional slow wind at the 
high latitudes during periods of solar maximum due to the latitudinal profile of the solar wind, which 
further contributes slow wind in the high latitude heliosheath. Because we are not using a time 
dependent solar wind which includes slow solar wind originating at high latitudes as 
observed during solar maximum, we overpredict the ENA flux within the high latitude lobes. 

We underestimate the ENA flux within the low latitude nose and tail for Cases 1 and 2. In these cases,
the solar wind dynamic pressure at 1 AU at the equator is 1.9 nPa. The \textit{IBEX} GDF data 
presented here is averaged over the years 2009 through 2013. For ENAs observed at the nose, it takes 
approximately two years from the time the solar wind is observed at 1 AU to cross the termination 
shock, charge exchange with an interstellar neutral to create an ENA, and then for the ENA to be 
observed by \textit{IBEX}. In the direction of the nose, IBEX is observing ENAs originating from solar
wind observed at 1 AU during the years 2007 through 2011. During this time, the solar wind dynamic 
pressure at 1 AU varied from 2.3 nPa, decreased to 1.2 nPa, and increased again to 2.2 nPa 
(Figure \ref{bcs}). The thickness of the heliosheath in the heliotail is greater than the 
thickness in the nose, but the distance over which ENAs are observed is limited by the extinction of 
parent PUIs. Based on the travel time estimations from \citet{Schwadron18} and assuming a cooling 
length of 100 AU in the heliotail, we can approximate the time delay for a 1.1 keV ENA in the tail to 
be approximately two to eight years. Therefore, \textit{IBEX} is observing in the tail ENAs 
originating from solar wind observed at 1 AU during the years 2001 through 2011. As noted in Figure 
\ref{bcs}, with the exception of the end of the year 2003 and the beginning of the year 2004, the 
dynamic pressure in the year 2008 displayed the lowest dynamic pressure as compared to the seven years
preceding it. The dynamic pressure affects the heating of the solar wind plasma crossing into the 
heliosheath, which affects ENA production. At the termination shock, the kinetic energy of the
solar wind is converted into thermal energy, so a higher dynamic pressure leads to a hotter PUI 
population. Therefore, our underprediction of ENA flux at the low latitudes may be attributable to our
exclusion of a time dependent solar wind, which would account for changes in dynamic pressure. 

\begin{figure*}[t!]
\centering
  \includegraphics[scale=0.63]{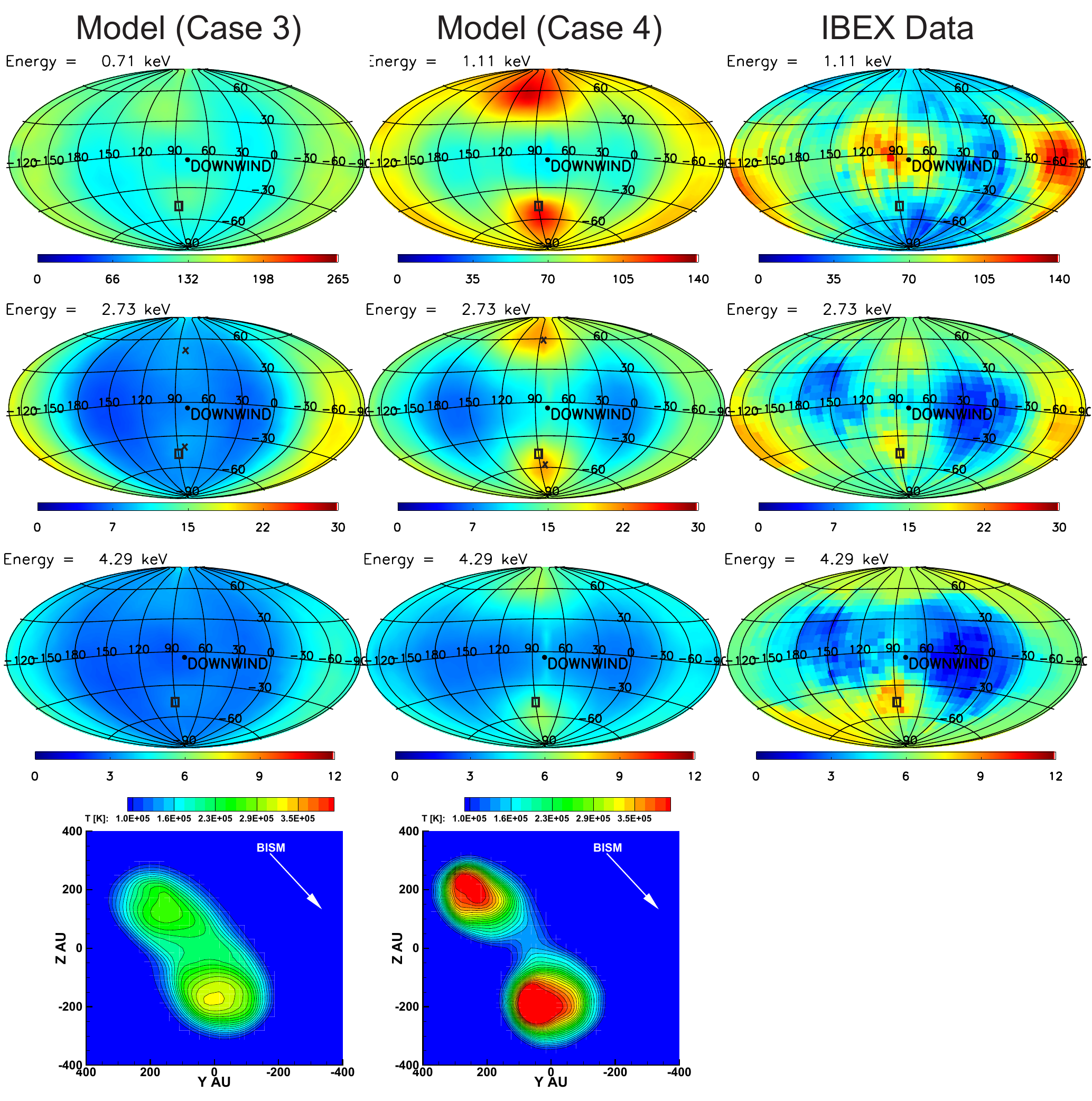}
  \caption{ENA sky maps displaying flux in units of (cm$^{2}$ s sr keV)$^{-1}$ 
  centered on the downwind (tail) direction. Simulated sky maps for the uniform 
  solar wind model with the radial component of the solar magnetic field, corresponding
  to the year 2008, set to be 2.94 nT at 1 AU (Case 3, left) and 6.45 nT at 
  1 AU corresponding to solar maximum-like conditions used in \citet{Opher15} 
  (Case 4, middle). Also included are \textit{IBEX} ENA maps from the first five 
  years (2009-2013) of observations \citep{Schwadron14}. Cuts for Case 3 (left) and Case 4 (middle) at 
  $x=600$ AU downtail viewed in the direction of ISM flow, with contours and lines of temperature. One can see 
  that the lobes are stronger   and exhibit a greater separation along the direction of the 
  interstellar magnetic   field in Case 4 than Case 3. Simulated sky maps are multiplied by a 
  factor of 1.8. The black box on each map corresponds to the location in the southern lobe 
  where the spectral indices are derived in Table \ref{spec}. For Case 3 in the 2.73 keV energy band the maximum
  flux within the lobes are located at ($\lambda$, $\beta$) = (84$^{\circ}$, 48$^{\circ}$) for the northern lobe and ($\lambda$, $\beta$) = (84$^{\circ}$, -36$^{\circ}$) for the southern lobe. For Case 4, the northern 
  lobe displays maximum flux at ($\lambda$, $\beta$) = (84$^{\circ}$, 60$^{\circ}$) and the 
  southern lobe has peak flux centered at ($\lambda$, $\beta$) = (84$^{\circ}$, -54$^{\circ}$) for the
  2.73 keV energy band. The location of maximum flux within the lobes of Cases 3 and 4 are marked by a 
  black `x'. }
  \label{uni}
\end{figure*}

The underprediction of ENA flux in the low latitude tail could also be attributable to not including 
the turbulent mixing between the interstellar and solar wind plasma in the tail along reconnected 
field lines. By not including this region of mixed material between HS and ISM – what we call ``open 
heliosheath", we are limiting our integration distance down the heliotail to where the magnetic field lines are
strictly solar. In contrast to the collimated lobes of solar wind plasma in the high latitude tail which extend
as far out as 600 AU from the Sun, the region of the low latitude heliotail where the field lines are 
strictly solar extends to approximately 350 AU in our model. As \citet{Kornbleuth18} discussed, PUIs 
within this region of ``open heliosheath” could be energized due to magnetic reconnection between the 
solar and interstellar magnetic field, which could possibly lead to an enhancement of ENAs created 
from this region. While our exclusion of this region could lead to a potential diminishment of ENAs in
the low latitude tail, the investigation of this region and its effect on ENAs will be left to a 
future study.

\begin{table*}[t!]
\centering
\begin{tabular}{cccccccccccc}
\tableline
 & & \textit{IBEX} Data & & Case 1 & & Case 2 & & Case 3 & & Case 4 \\
\tableline
$\gamma_{1}$ && 1.42 $\pm$ 0.16 && 0.64 && 0.52 && 1.75 && 1.49\\
$\gamma_{2}$ && 1.54 $\pm$ 0.12 && 1.73 && 1.58 && 2.32 && 2.39\\
\tableline
\end{tabular}
\caption{Spectral slopes for \textit{IBEX-Hi} data and Models corresponding to ENA flux 
modeled through the southern lobe at an ecliptic longitude of 90$^{\circ}$ and an 
ecliptic latitude of -42$^{\circ}$ (small black box on maps in Figures \ref{tail} and \ref{uni}). $\gamma_{1}$ is the spectal slope over the low-energy portion of the spectrum, 0.71-1.74 keV. $\gamma_{2}$ is the spectal slope over the high-energy portion of the spectrum, 1.74-4.29 keV.}
\label{spec}
\end{table*}

In Figure \ref{uni}, we present a comparison of Cases 3 and 4 along with \textit{IBEX} observations. 
Case 3 reflects the solar magnetic field of solar minimum conditions corresponding to 
Case 2. Case 4 uses a stronger solar magnetic field corresponding to conditions 
used in \citet{Opher15} and \citet{Kornbleuth18} reproducing a solar magnetic field strength in solar 
maximum-like conditions. When using a stronger solar magnetic field, the collimation of the solar wind
plasma is more pronounced. The additional heating leads to more ENA production. 
For Cases 1 and 2, the solar wind dynamic pressure is 1.9 nPa at 1 AU. For Cases 3 and 4, the
solar wind dynamic pressure is 2.7 nPa at 1 AU. The greater solar wind dynamic pressure in Cases 3 and
4 results in the outward motion of the termination shock, as compared to Cases 1 and 2; however, the 
greater magnetic field in Case 4 moves the termination shock inward. The increased solar magnetic 
field strength in Case 4 causes the termination shock to move 6 AU and 27 AU inwards relative to Case 
3 in the nose and tail directions, respectively. The closer termination shock results in denser 
solar wind plasma downstream of the termination shock in all directions, which enhances
ENA production. We conclude that the solar magnetic field is responsible for the ENA 
enhancement in the tail region in Figure \ref{uni}. While it has been suggested that 
the structure of the slow and fast solar wind result in the high latitude lobes observed by 
\textit{IBEX} at energies $>$ 2 keV, our results show that the collimation of the solar wind plasma by
the solar magnetic field acts as a critical contributor to the additional ENA flux enhancement and the
shape of the observed high latitude lobes.

The maximum flux within the lobes for the 2.73 keV energy band are located at ($\lambda$, $\beta$) = 
(84$^{\circ}$, 48$^{\circ}$) for the northern lobe and ($\lambda$, $\beta$) = (84$^{\circ}$, 
-36$^{\circ}$) for the southern lobe in Case 3. For Case 4, the northern 
lobe displays maximum flux at ($\lambda$, $\beta$) = (84$^{\circ}$, 60$^{\circ}$) and the 
southern lobe has peak flux centered at ($\lambda$, $\beta$) = (84$^{\circ}$, -54$^{\circ}$) for the
2.73 keV energy band. Figure \ref{uni} also shows cuts down the heliotail at $x=600$ AU with contours of
temperature. As the solar magnetic field increases, the lobes move to higher latitude. This is seen 
both in the ENA maps of Figure \ref{uni} and the $x=600$ AU cuts. This is because the heliospheric 
jets with a stronger solar magnetic field are more resistant to the pressure from the ISM. We also 
find in Figure \ref{uni} that the lobes in both models are oriented along the direction of the 
interstellar magnetic field, which is noted for Cases 1 and 2 as well. 

In Table \ref{spec}, we present the spectra for the modeled ENA flux in the 
southern lobe corresponding to an ecliptic longitude and latitude of 
($\lambda$, $\beta$) = (90$^{\circ}$, -42$^{\circ}$) as compared to \textit{IBEX} data. The location is 
represented by a black box on the ENA maps in Figures \ref{tail} and \ref{uni}. We only consider the
southern lobe because the northern lobe observed by \textit{IBEX} crosses the path of the 
\textit{IBEX} Ribbon, while the southern lobe does not. The spectral slope 
represents how the flux changes as a function of energy for a given latitude and 
longitude. The flux can be approximated by $J \propto E^{-\gamma}$. Previous 
studies have shown that there is both a low-energy and high-energy component to 
the polar ENA spectra that comprises two distinct spectral slopes. 
\citet{Dayeh11, Dayeh14} examined these two slopes by dividing the spectrum into 
two portions at 1.74 keV. We divide our spectra in two bins as in 
\citet{Dayeh11}, with $\gamma_{1}$ corresponding to the flux for the 0.71, 1.11, 
and 1.74 keV energy bands, and $\gamma_{2}$ corresponding to the flux for the 
1.74, 2.73, and 4.29 keV energy bands. The spectrum of $\gamma_{1}$ reflects 
the signature of the slow solar wind, while $\gamma_{2}$ reflects the 
signature of the fast solar wind. 

We find that $\gamma_{1}$ is 0.52 and 0.64 for Cases 1 and 2, 
respectively. These values are considerably lower than what is seen in \textit{IBEX} 
observations ($\gamma_{1}\sim$ 1.42 $\pm0.16$). As mentioned previously, our model does not 
include a time dependent solar wind model. As the solar cycle progresses, \textit{IBEX} is 
observing ENAs produced during both solar minimum and solar maximum conditions. 
We only are modeling solar minimum conditions and are therefore excluding slow wind at the poles as 
would be seen during solar maximum, so it is reasonable to expect that we would not be able to 
reproduce \textit{IBEX} spectra for $\gamma_{1}$. For Cases 3 and 4 we find steeper spectral 
slopes ($\gamma_{1}\sim$ 1.75 and $\gamma_{1}\sim$ 1.49, respectively) than is 
present in the \textit{IBEX} data, but the spectrum is closer in approximating the \textit{IBEX} 
data than Cases 1 and 2 due to the slow solar wind.

We find that $\gamma_{2}$ is 1.58 and 1.73 for Cases 1 and 2, respectively. Case 
2 has a flatter spectrum in the pole, as compared to Case 1, which is a result of 
the heating caused by the solar wind collimation by the solar magnetic field 
present in Case 2. Since our poles are populated with fast solar wind due to the 
solar minimum conditions, we find that $\gamma_{2}$ for our model is in good 
agreement with \textit{IBEX} observations ($\gamma_{2}\sim$ 1.54 $\pm0.12$), with Case 2 having 
better agreement due to the presence of the solar magnetic field. Cases 3 and 4 
have spectra of $\gamma_{2}\sim$ 2.32 and $\gamma_{2}\sim$ 2.39, respectively, 
which yield significantly steeper slopes compared to the \textit{IBEX} data due to 
the lack of fast solar wind present in the poles for this particular model.

\section{Discussion}
The ``croissant-like" heliosphere with a latitudinally dependent solar wind is able 
to qualitatively reproduce the \textit{IBEX} maps from the first five years of observations. 
A more quantitative comparison with a fully time dependent solar wind is left for a 
future study. 

We find that the solar magnetic field plays a critical role in the 
observed ENA structure. The combination of the slow and fast profile of the solar 
wind coupled with the collimation of the solar wind plasma by the solar magnetic 
field leads to the high latitude lobes observed by \textit{IBEX} at energies $>$ 2 keV. As 
noted in Figure \ref{tail}, while the latitudinal variation of the solar wind can 
play a role in the creation of high latitude lobes within the tail, the influence 
of the solar magnetic field leads to an enhancement of ENA flux within these 
lobes. This is the first work to explore the effect of the changing solar magnetic field 
strength on ENA maps.

We predict that the high latitude lobes of strong ENA flux will remain during the 
course of the solar cycle. As the solar cycle progresses towards solar maximum, 
the solar wind structure will become more uniform, but the magnitude of the solar 
magnetic field will increase in strength. Therefore, the collimation of the solar 
wind plasma by the solar magnetic field strengthens during solar maximum, 
leading to the persistence of high latitude lobes as shown in \citet{Kornbleuth18}.
Our results suggest that the high latitude lobes observed by \textit{IBEX} from 2009 to 2017 
\citep{Schwadron18} are the result of the collimation of the solar wind plasma by the magnetic field 
and the increase of the magnetic field during solar maximum.

\acknowledgments
The authors would like to acknowledge Dr. Nathan 
Schwadron for supplying information on IBEX GDF 
uncertainties and to acknowledge helpful
discussions with Drs. Maher Dayeh and Dan Clemens.
This work was supported by NASA Headquarters under
the NASA Earth and Space Science Fellowship Program, 
Grant 80NSSC18K1202. M.O. was partially supported by 
NASA grant 18-DRIVE18{\_}2-0029, Our Heliospheric 
Shield, 80NSSC20K0603. Resources supporting this work 
were provided by the NASA High-End Computing (HEC) 
program through the NASA Advanced Supercomputing (NAS) 
Division at Ames Research Center. The authors would 
like to thank the staff at NASA Ames Research Center 
for the use of the Pleiades super-computer under the 
award SMD-18-1875. JMS acknowledges support from the 
Polish National Agency for Academic Exchange (NAWA) 
Bekker Program Fellowship PPN/BEK/2018/1/00049.


\end{document}